\documentstyle[preprint,aps]{revtex}
\begin{document}
\draft
%\flushbottom
\preprint{\vbox{
\hbox{IFT-P.043/2002}
%\hbox{physics/0207012 }
\hbox{July 2002}
}}
\tightenlines
\title{Schwinger's oscillator method, supersymmetric quantum mechanics
and massless particles~\footnote{Publicado em Revista Brasileira de Ensino de
F\'\i sica {\bf24}, 41 (2002).}}
\author{F. M. Mej\'\i a$^1$ and V. Pleitez$^2$ }
\address{
$^1$Escuela de F\'\i sica\\ Facultad de Ciencias Naturales y
Matem\'atica\\ Universidad de El Salvador\\ El Salvador\\
$^2$Instituto de F\'\i sica Te\'orica\\
Universidade Estadual Paulista\\
Rua Pamplona 145\\ 
01405-900-- S\~ao Paulo, SP\\
Brazil}
\maketitle
\begin{abstract}

We consider the Schwinger's method of angular momentum addition
using the $SU(2)$ algebra with both
a fermionic and a bosonic oscillator. 
We show that the total spin states obtained are: one boson singlet
state and an arbitrary number of spin-1/2 states, the later ones are energy 
degenerate. It means that we have in this case supersymmetric quantum 
mechanics and also the addition of angular momentum for massless particles. 
We review too the cases of two bosonic and fermionic oscillators.
\end{abstract}
\narrowtext

\section{Introduction}
\label{sec:intro}

The usual method for defining the angular momentum in quantum mechanics is by
means of the commutation relations satisfied by its components $J_i,\;i=x,y,z$;
and  by solving the eigenvalue problem for $\vec J^2$ and $J_z$ assuming that
the components $J_i$ are observables. {}From this, 
the allowed values for the eigenvalues of $\vec J^2$ 
and $J_z$, denoted $j$ and $m$, respectively, are obtained. They run over 
the values: $j=0,1/2,1,3/2,...$ and $-j\leq m\leq j$~\cite{am}. 
In this case the angular momentum operators $J_i$ are the infinitesimal 
generators for the $SO(3)\sim SU(2)$ algebra. 
The relation between spin and $SU(2)$ symmetry is 
maintained in relativistic field theory since the little group for massive 
particles is just the rotation group~\cite{ew,sw}.
For massive spin-$j$ particles we can always go to the rest frame, thus their
spin degrees of freedom transform according to a $(2j+1)$-dimensional 
representation of $SU(2)$, that is, we have $2j+1$ polarization states.

In the case of massless particles it is not possible to go to the rest frame, 
so the spin is no longer 
described by $SU(2)$. In this case the little group is the Euclidean 
$ISO(2)$, denoted also by $E(2)$. This group consists of rotations by an
angle $\phi$  around the $z$-direction (assuming this as the direction 
of the motion) and translations in the Euclidean 
plane perpendicular to the axis $z$. Its irreducible representations 
must either represent the
translation by zero, or be infinite-dimensional. If $\chi$ is an eigenvector
of the translation generators, $e^{i\phi J_z}\chi$ will be also an eigenvector
rotated by an angle $\phi$. There is no room in physics besides the momentum for
another continuous quantum number, so physical massless particles correspond 
only to the first kind of representations (i.e., the trivial ones)~\cite{sw}. 
This leaves only $J_z$ as symmetry operator, so the physical representations of
$E(2)$ are one-dimensional and labeled by the helicity $\lambda$. 
$J_z\vert\lambda\rangle=\lambda\vert\lambda\rangle$. 
This is why the polarization states of a massless 
particle with spin $j$ are only $\pm j$. We can see this by considering the 
second Casimir invariant $W^\mu W_\mu=-M^2j(j+1)$ (the other is $p^2=M^2>0$) 
where $W^\mu$ is the Pauli-Lubanski pseudovector defined as
\begin{equation}
W_\mu=-\frac{1}{2}\varepsilon_{\mu\nu\rho\sigma}
J^{\nu\rho}P^\sigma,
\label{pauli}
\end{equation}
with $J^{\nu\rho}$ and $P^\sigma$ denoting the generators of the Poincar\'e
group; $\varepsilon_{\mu\nu\rho\sigma}$ is the totally antisymmetric 
symbol in four dimensions. Since $M^2=0$ we have for a state of the four 
momentum $k$,
\begin{equation}
W\cdot W\vert k\rangle=0,\quad P\cdot P\vert k\rangle=0
\label{con}
\end{equation}
and, since $W^\mu P_\mu=0$,
\begin{equation}
W\cdot P\vert k\rangle=0.
\label{con2}
\end{equation}
So, $W^\mu$ and $P^\mu$ are orthogonal and both lightlike. This means that 
they must be proportional 
\begin{equation}
(W^\mu-\lambda P^\mu)\vert k\rangle=0,
\label{con3}
\end{equation}
and we have the result that the state of a massless particle is characterized 
by one number $\lambda$, which is the ratio of $W^\mu$ and $P^\mu$ and so it
has the dimension of angular momentum. It is called, as we said before, 
helicity. If parity is included the helicity takes on two values, $\lambda$ 
and $-\lambda$. The fact that $\lambda$ can be integral or half-integral is 
due to the fact that $e^{i4\pi\lambda}$ must be unity, and hence $\lambda$ must 
be integer or half-integer~\cite{sw}.

Several years ago Schwinger worked out a connection between the algebra of 
angular-momentum and two uncoupled bosonic oscillators~\cite{js1}. 
The Schwinger's scheme permits to calculate the eigenvalues $j$ and $m$, in 
terms of the operator number of the uncoupled oscillators. 
The results agree with the general method for massive particles discussed 
above. 
The case of massless particles, however, does not arise within 
the Schwinger's scheme. So, the method must be generalized. 
In this work we take a first step to a more complete generalization 
by considering one or more fermionic oscillators. 

The outline of this work is the following. The Schwinger's method which consists
of two bosonic oscillators ($BB$) is reviewed in Sec.\ref{sec:bb}. Next,  we
generalize the scheme by considering i) two fermionic oscillators $(FF)$ 
in Sec.~\ref{sec:ff} and, ii) one bosonic and one fermionic oscillator
$(FB)$ in Sec.~\ref{sec:fb}. In the last cases only spin zero and spin 1/2 are
generated. In the last section we show that in the BF case the spin 1/2
particles are infinitely degenerated and supersymmetric quantum mechanics
naturally arises.

\section{Two bosonic oscillators (BB)}
\label{sec:bb}

In this section we will briefly review the Schwinger scheme
~\cite{js1,sakurai} by considering two simple bosonic oscillators with 
annihilation and creation operators $a_i$ and $a_i^\dagger,\;i=1,2$, 
respectively. The number operators are (throughout this work we will use 
$\hbar=1$) $N_i\equiv a_i^\dagger a_i$ and assuming the commutation relations
$[a_i,a_j^\dagger]=\delta_{ij}$, it follows that 
\begin{equation}
[N_i,a_j]=-a_i\delta_{ij},\quad [N_i,a_j^\dagger]=a_i^\dagger\delta_{ij},
\qquad \mbox{(no summation)}.
\label{3}
\end{equation}
We also assume that another pair of operators of the same oscillator or 
of different oscillators commute. It means that the two oscillators are 
uncoupled.
Because $N_1$ and $N_2$ commute, we can build up simultaneous eigenstates
of $N_1$ and $N_2$ with eigenvalues $n_1$ and $n_2$, respectively. 

Next, we define 
\begin{mathletters}
\label{def1}       %7
\begin{equation}
J_+\equiv a_1^\dagger a_2,\qquad J_-\equiv a_2^\dagger a_1,
\label{7a}
\end{equation}
and
\begin{equation}
J_z\equiv \frac{1}{2}(a_1^\dagger a_1-a^\dagger_1 a_2)=\frac{1}{2}(N_1-N_2).
\label{7b}
\end{equation}
\end{mathletters}
These operators satisfy the $SU(2)$ commutation relations
\begin{equation}
[J_z,J_{\pm}]=\pm J_{\pm},\qquad [J_+,J_-]=2J_z.
\label{su2}%8
\end{equation}

Defining the total number operator $N$ (with eigenvalues $n_1+n_2$)
\begin{equation}
N\equiv N_1+N_2=a_1^\dagger a_1+a^\dagger_2 a_2,
\label{9}
\end{equation}
it follows that the quadratic Casimir operator
\begin{equation}
\vec J^2=J^2_z+\frac{1}{2}\left(J_+J_-+J_-J_+\right),
\label{10}
\end{equation}
can be written as
\begin{equation}
\vec J^2=\frac{N}{2}\left(\frac{N}{2}+1 \right).
\label{11}
\end{equation}

If we associate spin up $(m=1/2)$ with one
quantum unit of the $N_1$ oscillator and spin down $(m=-1/2)$ with one 
quantum unit of the $N_2$ oscillator,
it is possible to imagine a spin $1/2$ ``particle'' with  spin up (down)
with each quantum unit of the $N_1(N_2)$ oscillator. The eigenvalues $n_1$
and $n_2$ are just the number of spin up and spin down ``particles'', 
respectively.
We will see that the association of half-integral spin with bosonic 
oscillators is necessary, if we want to construct a general 
$\vert j,m\rangle$ state with $j=0,1/2,1,3/2,2,...$ and $-j\leq m\leq j$.

Turning back to the $J_{\pm}$ operators defined in Eqs.~(\ref{def1}), we
see that $J_+$ destroys one unit of spin down with the 
$z$-component of angular $-1/2$ and creates one unit of spin up with the
$z$-component of angular momentum $+1/2$. So, the $z$-component of 
angular momentum is therefore increased by $1$. Likewise $J_-$ destroys
one unit of spin up and creates one unit of spin down, the $z$-component
of angular momentum is therefore decreased by $1$.
As for the $J_z$ operator, it simply counts $1/2$ ($\hbar=1$) times the 
difference between $n_1$ and $n_2$, just gives the $z-$component of the total
angular momentum. Hence, the action of the $J_\pm$ and $J_z$ operators on the
eigenstates of the $\vert n_1,n_2\rangle$ is given by 
\begin{mathletters}
\label{12}
\begin{equation}
J_+\vert n_1,n_2\rangle=a_1^\dagger a_2\vert n_1,n_2\rangle=
[n_2(n_1+1)]^{1/2}\;\vert n_1+1,n_2-1\rangle,
\label{12a}
\end{equation}
\begin{equation}
J_-\vert n_1,n_2\rangle=a_2^\dagger a_1\vert n_1,n_2\rangle=
[n_1(n_2+1)]^{1/2}\;\vert n_1-1,n_2+1\rangle,
\label{12b}
\end{equation}
\begin{equation}
J_z\vert n_1,n_2\rangle=\frac{1}{2}(N_1-N_2)\vert n_1,n_2\rangle=
\frac{1}{2}(n_1-n_2)\vert n_1,n_2\rangle.
\label{12c}
\end{equation}
\end{mathletters}
Notice that, the sum $n_1+n_2$ which gives the total number of spin $1/2$ 
particles remains unchanged.
If we substitute
\begin{equation}
n_1\to j+m,\qquad n_2\to j-m,
\label{13}
\end{equation}
Eqs.~(\ref{12}) reduce to the usual factors
\begin{mathletters}
\label{14}
\begin{equation}
J_+\vert n_1,n_2\rangle=[(j-m)(j+m+1)]^{1/2}\,\vert j+m+1,j-m-1\rangle,
\label{14a}
\end{equation}
\begin{equation}
J_-\vert n_1,n_2\rangle=[(j+m)(j-m+1)]^{1/2}\,\vert j+m-1,j-m+1\rangle,
\label{14b}
\end{equation}
\begin{equation}
J_z\vert j+m,j-m\rangle=m\vert j+m,j-m\rangle,
\label{14c}
\end{equation}
\end{mathletters}
and the eigenvalues of the quadratic Casimir operator $\vec{J}^2$ defined
in Eq.~(\ref{11}) become
\begin{eqnarray}
\vec J^2\vert j+m,j-m\rangle&=&\frac{n_1+n_2}{2}\left[\frac{n_1+n_2}{2}+1 
\right] \vert j+m,j-m\rangle\nonumber \\ &\equiv& j(j+1)\vert j+m,j-m\rangle.
\label{15}
\end{eqnarray}

The connection between the oscillator matrix elements and angular momentum 
matrix elements can be seen by defining
\begin{equation}
j\equiv \frac{n_1+n_2}{2},\qquad m\equiv \frac{n_1-n_2}{2},
\label{16}
\end{equation}
in place of $n_1$ and $n_2$ to characterize simultaneous eigenkets of 
$\vec{J}^2$ and $J_z$. Hence, the most general $N_1,N_2$ eigenket is
\begin{equation}
\vert j,m\rangle=\frac{(a_1^\dagger)^{j+m} (a_2^\dagger)^{j-m}}
{[(j+m)!(j-m)!]^{1/2}}\,\vert 0,0\rangle.
\label{17}
\end{equation}
If $j=m$ we have the largest eigenvalue for $J_z$
\begin{equation}
\vert j,j\rangle=\frac{(a_1^\dagger)^{2j}}{[(2j)!]^{1/2}}\,\vert 0,0\rangle,
\label{18}
\end{equation}
so we can imagine this state to be build up of $2j$ spin $1/2$ particles with
their spin all pointing in the positive $z$-direction.  Hence, as we said
before, in this scheme an object of high $j$ can be visualized as being
made up of primitive spin $1/2$ ``particles'', $j+m$ of them with spin up and
the remaining $j-m$ of them with spin down. This of course, does not mean
that an object of angular momentum $j$ is a composite system of spin $1/2$
particles. It means only that, as far as the transformation properties are
concerned, we can visualized any object of arbitrary angular momentum
$j$ as a composite system of $2j$ spin $1/2$ particles formed in the manner
indicated by Eq.~(\ref{17}). This is the well known Schwinger's oscillator
method~\cite{sakurai}. 

\section{Two fermionic oscillators (FF)}
\label{sec:ff}

Let us consider two fermionic oscillators with annihilation and creation
operators denoted by $F_i$ and $F_i^\dagger$, $i=1,2$. Then
\begin{equation}
         \{F_i,F_j^\dagger\}=\delta_{ij},
\label{19}
\end{equation}
and any other pair of operators anticommuting. The number operators are defined
as usual $N_i=F^\dagger_iF_i$, $i=1$ or $2$, and they satisfy
\begin{equation}
[N_i,F_j]=-F_i\,\delta_{ij},\qquad
[N_i,F_j^\dagger]=F_i^\dagger\,\delta_{ij},
\qquad\mbox{(no summation)}.
\label{20}    
\end{equation}

However, from Eq.~(\ref{19}) it follows that
\begin{equation}
N_i(N_i-1)=0,\qquad i=1 \;\;\mbox{or} \;\; 2;
\label{21}
\end{equation}
so, the only eigenvalues of $N_i$, denoted by $n_i$, are $0$ or $1$. The total 
number operator $N=\sum_iN_i$, has eigenvalues $0,1$ or $2$.

As in the case of two bosonic oscillators, we can construct simultaneous 
eigenkets of $N_1$ and $N_2$. Eqs.~(\ref{17}) are valid but with 
the substitution $a_i\to F_i$, $a_i^\dagger\to F_i^\dagger$ and with the 
constraint upon $n_1,n_2$ given above. Thus as in Eqs.~(\ref{def1}) we can
define 
\begin{mathletters}
\label{def2}
\begin{equation}
J_+\equiv F^\dagger_1F_2,\qquad J_-\equiv F_2^\dagger F_1,
\label{22a}
\end{equation}
\begin{equation}
J_z\equiv\frac{1}{2}\left(F^\dagger_1 F_1-F_2^\dagger F_2 \right)
=\frac{1}{2}\left(N_1-N_2\right),
\label{22b}
\end{equation}
\end{mathletters}
which satisfy, as can be easily verified, the $SU(2)$ commutation relations in 
Eq.~(\ref{su2}). {}From the point of view of the $SU(2)$ algebra both 
cases, two bosonic oscillators and two fermionic oscillators are equivalent.
Notwithstanding $\vec{J}^2$, defined in Eq.~(\ref{10}), when written in
terms of the number operators, instead of Eq.~(\ref{11}) is given by
\begin{equation}
\vec J^2=\frac{N}{2}\left(\frac{N}{2}+1\right)-2N_1N_2.
\label{23}
\end{equation}
Since the eigenvalues of $N_1$ and $N_2$ can assume only the values $0$ or 
$1$, we see from Eq.~(\ref{23}) that the respective eigenvalues for 
$\vec{J}^2$ are $0$ and $3/4$. If we interpret these values in the form 
$j(j+1)$, $j\geq 0$, we see that only $j=0$ and $j=1/2$ are allowed. The 
eigenvalues of $J_z$ defined in Eq.~(\ref{22b}) are $0,1/2,-1/2$~\cite{lerda}:
\begin{mathletters}
\label{24}
\begin{equation}
J_z\vert0,0\rangle=0, \quad J_zF_1^\dagger F_2^\dagger\vert0,0\rangle=0,
\quad J_z\vert1,1\rangle=0,\quad J^2\vert1,1,\rangle=0,
\label{24a}
\end{equation}
\begin{equation}
J_z F_1^\dagger\vert0,0\rangle=\frac{1}{2} F_1^\dagger\vert0,0\rangle,
\quad J_z F_2^\dagger\vert0,0\rangle=-\frac{1}{2} F_2^\dagger\vert0,0\rangle.
\label{24b}
\end{equation}
\end{mathletters}
Hence, it seems that with two fermionic oscillators we can build up only
one spin-$1/2$ ($F^\dagger_1\vert 0,0\rangle,\vert F^\dagger\vert 0,0\rangle$)
and two spin-$0$ states ($\vert0,0\rangle,\vert1,1\rangle$). In others words,
although the system also satisfies the usual angular-momentum commutation
relations in Eq.~(\ref{su2}), only
these two values for the total angular-momentum are allowed. We call this
situation a {\it constrained realization} of the $SU(2)$ algebra.

If we associate spin up $(m=1/2)$ with one
quantum unit of the $F_1$ oscillator and spin down $(m=-1/2)$ with one 
quantum unit of the $F_2$ oscillator,
it is possible to imagine one spin $1/2$ ``particle'' with  spin up (down)
with each quantum unit of the $F_1(F_2)$ oscillator. 
As in Sec.~\ref{sec:bb} 
the spins are along the $z-$direction and the eigenvalues 
$n_1$ and $n_2$ are just the numbers of spins up and spins down, respectively. 
However, in the present case if $n_1=n_2=0$ the total spin is also zero;
if $n_1=n_2=1$ both spins are in opposite direction and the total spin 
vanishes again. On the other hand, if $n_1=1,n_2=0$ the total spin is $1/2$
and the projection on the $z-$axis is $1/2$; if $n_1=0,n_2=1$ the total spin
is again $1/2$ but its projection in that axis is $-1/2$. Notice, however,
that this case does not correspond neither to the massive nor to the 
massless particle cases. It can be applied to both kind of particles.

\section{One fermionic and one bosonic oscillators (FB)}
\label{sec:fb}

We have seen that both, the usual angular momentum addition and the Schwinger's
scheme are valid for the case of massive particles~\cite{sakurai}. 
The case of massless particles, however, does not arise 
neither within the usual approach nor in the Schwinger's scheme. 
So, the method must be generalized. In this work we take a first step to 
get a more complete generalization by considering one fermionic 
oscillator. 

The interesting feature of the 
Schwinger's scheme is that it allows us to obtain what are the values of the 
weights or roots that are realized in the $SU(2)$ algebra. For instance, in
the original work of Schwinger, all representations of the $SU(2)$ algebra
arise and it is exactly equivalent, as we  said before, to the theory of 
the angular momentum addition. 
This is however a consequence of the bosonic nature of the 
oscillators. Notwithstanding, when both oscillators are fermionic, 
although the algebraic relations are still valid the method does not 
coincide with the usual addition of angular momentum in the sense that only 
restricted values for the eigenvalues of the angular momentum operator
are allowed: only two spinless states and one spin-1/2
state are obtained and it is impossible to recover 
the full set of the unitary representation of $SU(2)$. 
This is a consequence of the fermionic character of the 
operators with which we implement the realization of the $SU(2)$ 
algebra~\cite{lerda}. 
On the other hand, when one of the oscillators is bosonic and the other one is  
fermionic, a usual $SU(2)$ algebra is still realized but also in a restricted 
sense. This is the case that we will consider here.

Let us consider the case of two oscillators, one of them a bosonic 
oscillator ($a,a^\dagger $) and the other one a fermionic oscillator
($F,F^\dagger $). It means that
\begin{equation}
\left[a,a^\dagger \right]=1,\qquad \left\{F,F^\dagger \right\}=1,
\label{25}
\end{equation}
and any pair of operators commutes if both of them are  bosonic operators or,
if one of them is a bosonic operator and the other is a fermionic operator;
they anticommute if both of them are fermionic operators. 

As before, we will use the following notation: 
The number operators are denoted by $N_B=a^\dagger a$, $N_F=F^\dagger F$ 
and $N=N_B+N_F$, with eigenvalues $n_B,n_F$ and $n=n_B+n_F$, respectively.
Let us define
\begin{mathletters}
\label{def}
\begin{equation}
J_+\equiv a^\dagger F\,(N_B+1)^{-1/2},\qquad J_-\equiv 
(N_B+1)^{-1/2}\,F^\dagger a,
\label{26a}
\end{equation}
\begin{equation}
J_z\equiv\frac{1}{2}\,\left[a^\dagger(N_B+1)^{-1} a(1-N_F)-
N_F\right].
\label{26b}
\end{equation}
\end{mathletters}

If the simultaneous eigenkets of $N_B$ and $N_F$ are denoted by 
$\vert n_B,n_F\rangle$ when necessary we will use the closure relation 
$\sum\vert n_B,n_F \rangle \langle n_B,n_F\vert={\bf1}$ in order to get
a result that it is not state dependent. For instance
\begin{equation}
(N_B+1)^{-1}
\sum_{n_B,n_F}\vert n_B,n_F \rangle \langle n_B,n_F\vert=
 \sum_{n_B,n_F}(n_B+1)^{-1}\vert n_B,n_F \rangle \langle n_B,n_F\vert,
\label{cr}
\end{equation}
and similarly for $(N_B+1)^{-1/2}$.
Then, it is possible to verify that the operators defined in Eq.~(\ref{def}) 
satisfy the commutation relations of the $SU(2)$ algebra given
in Eq.~(\ref{su2}) and that
\begin{equation}
J_+\vert n_B,n_F\rangle=
a^\dagger (N_B+1)^{-1/2}F \,\vert n_B,n_F\rangle=
\sqrt{n_F}\,\vert n_B+1,n_F-1\rangle,
\label{29a}
\end{equation}
that is, $J_+\vert n_B,n_F\rangle=\vert n_B+1,0\rangle$
if $n_F=1$ and  $J_+\vert n_B,n_F\rangle=0$ if $n_F=0$. Similarly,
\begin{equation}
J_-\vert n_B,n_F\rangle=F^\dagger (N_B+1)^{-1/2}a\,\vert n_B,n_F\rangle=
\sqrt{1-n_F}\,\vert n_B-1,n_F+1\rangle,
\label{29b}
\end{equation}
hence $J_-\vert n_B,n_F\rangle=0$ when $n_F=1$ or $n_B=0$, and    
$J_-\vert n_B,n_F\rangle=\vert n_B-1,1\rangle $ when $n_F=0$.

Next, we obtain the quadratic Casimir operator 
\begin{equation}
\vec{J}^2=\frac{1}{2}\left(\frac{1}{2}+1 \right)\left[ 
a^\dagger (N_B+1)^{-1}a(1-N_F)+N_F\right],
\label{33}
\end{equation}
and the hamiltonian of the system can be written as
\begin{equation}
H=\{ J_+,J_-\}=a^\dagger(N_B+1)^{-1}a(1-N_F)+N_F,
\label{susy}
\end{equation}
and it satisfies $[H,J_\pm]=0$.

For the state $\vert n_B,n_F\rangle=\vert0,0\rangle$ we have
\begin{equation}
J_z\vert 0,0\rangle=0,\;J^2\vert 0,0\rangle=0,
\label{jz1}
\end{equation}
while for the states $\vert n_B,n_F\rangle\not=\vert 0,0\rangle$ it follows
\begin{equation}
J_z\vert n_B,n_F\rangle=\left(\frac{1}{2}-n_F\right)\vert n_B,n_F\rangle,
\label{jz2}
\end{equation}
and
\begin{equation}
J^2\vert n_B,n_F\rangle=\frac{1}{2}
\left(\frac{1}{2}+1\right)\vert n_B,n_F\rangle.
\label{j22}
\end{equation}

We see from Eqs.~(\ref{jz1}) and (\ref{j22}) that as in FF case, only spin 0 and
1/2 are generated. However an interesting difference appears 
when one of the oscillators is bosonic as we will see in the next section. 

\section{Conclusions}
\label{sec:con}

In the usual supersymmetric quantum mechanics,
the equality $\omega_B=\omega_F$ is imposed by hand~\cite{cooper}. 
In the Schwinger's scheme it is implicitly assumed that both oscillators
have the same frequency, $\omega$. In the case of both one fermionic and one 
bosonic oscillator this implies that $\omega_B=\omega_F$. It means that 
we have a symmetry of the combined bosonic and fermionic oscillators, 
that is, we have a supersymmetry. 
In fact, using Eq.~(\ref{susy}), we have
\begin{equation}
H\vert0,0\rangle=0,\quad H\vert n_B,n_F\rangle=\vert n_B,n_F\rangle.
\label{energia}
\end{equation}
So, we can identify $Q=J_+$ and 
$\bar Q=J_-$ as the supersymmetry generators. Note that, in fact, from 
Eqs.(\ref{29a}) and (\ref{29b}) we have
\begin{equation}
J_+\vert n_B,1\rangle=\vert n_B+1,0\rangle,\quad
J_-\vert n_B+1,0\rangle=\vert n_B,1\rangle.
\label{susy2}
\end{equation}
Then, the bosonic states $\vert n_B+1,0\rangle$ have the same energy than 
their fermionic partners $\vert n_B,1\rangle$. Only the vacuum state is not 
degenerate as it can be seen from Table I.

In fact, since in this case we have $N=1$ supersymmetric quantum mechanics, 
we can introduce a Grassmann parameter $\theta$ ($\theta^2=0$), and if we define
\begin{equation}
J_1=\theta(J_++J_-),\quad J_2=i\theta(J_+-J_-),
\label{ndef}
\end{equation}
we can verify that
\begin{mathletters}
\label{ob}
\begin{equation}
[J_z, J_1]=iJ_2,\quad 
[J_z, J_2]=-iJ_1
\label{ob1}
\end{equation}
and
\begin{equation}
[J_1,J_2]=-4i\theta^2 J_z=0.
\label{ob3}
\end{equation}
\end{mathletters}
This commutation relation defines the Euclidean group $E(2)$ and, as we
mentioned in Sec.~\ref{sec:intro}, 
it is well known that in the relativistic theory this is the little group 
related to massless particles~\cite{sw} and for this reason the polarization 
states of massless particles with spin $j$ are only $\pm j$. 
Thus we  can interpret our result as follows: 
since we are in a non-relativistic domain, i.e. just one
SU(2), only spinless
and spin one-half particles are allowed to be massless since in this
case both $2j+1$ or $\pm j$ degeneration coincide.
{}From the point of view of the angular momentum addition, a supersymmetric 
transformation
\begin{equation}
\vert n_B+1,0\rangle\Longleftrightarrow \vert n_B,1\rangle,
%\quad
%\mbox{or}\quad j\stackrel{parity}{\Longleftrightarrow} -j
\label{parity}
\end{equation}
is equivalent to a parity transformation which makes $j\to-j$ and 
supersymmetric quantum mechanics is e\-qui\-va\-lent to a ``constrained SU(2)
algebra''. 

If we consider $SU(2)\otimes SU(2)^\prime$, which corresponds to the 
relativistic case, we can have massless states with $j=0,1/2,1$ but 
not with $j>1$. 
A way to
overcome this problem is to consider a two-component spinor field as
in Ref.~\cite{lerda} as we will show elsewhere.
   
Finally, we would like to pointed out that it may be interesting to 
considered these extensions of the Schwinger scheme for $SU(n)$ and
also in relativistic field theories~\cite{js2}. 

\acknowledgements
This work was supported partially by Funda\c{c}\~ao de Amparo \`a Pesquisa
do Estado de S\~ao Paulo (FAPESP), Conselho Nacional de 
Ci\^encia e Tecnologia (CNPq) and by Programa de Apoio a
N\'ucleos de Excel\^encia (PRONEX). One of us (VP) would like to thank
J. F. Gomez for calling his attention to Refs.~\cite{lerda,cooper}.

%\begin{table}
%\caption { $SU(2)$ representations with the Schwinger's method with 
%two bosonic 
%oscillators. The allowed values of $m$ appear along the antidiagonal arrow.}
%\label{t1}
%\end{table}
%\begin{table}
%\caption {Same as in Table I. but with one bosonic and one fermionic 
%oscillators. }
%\label{t2}
%\end{table}

\newpage

\noindent Table I. States obtained by the Schwinger scheme and the
respective energies.

%\begin{table}
%\caption{...}
\begin{center}
\begin{tabular}{||c|c||c|c|c||}\hline 
$n_B$ & $n_F$  & $j$ & $m$  & $E$  \\ \hline 
0 & 0  & 0 & 0 & 0 \\
& 1 &1/2&-1/2&1\\
1 & 0 & 1/2 & 1/2 & 1 \\
 & 1  & 1/2 & -1/2 & 1 \\
2 & 0 & 1/2 & 1/2 & 1 \\
  & 1 & 1/2 & -1/2 & 1 \\
$\cdots$ & $\cdots$ & $\cdots$ & $\cdots$ & $\cdots$ \\ \hline
\end{tabular}
\end{center}
%\label{t1}
%\end{table}

%%%%%%%%%%%%%%%%%%%%%%%%%%%%%%%%%%%%%%%%%%%%%%%%%%%%%%%%%%%%%%%

\end{document}